\documentstyle[12pt]{article}

\makeatletter

\def\wgas@ver{2.2}
\def\wgas@pageid{\xdef\@thefnmark{\null}
\@footnotetext{This manuscript was prepared with the
		   AAS WGAS \LaTeX\ macros v\wgas@ver}}
\ifnum\@ptsize<2
\fi
\ps@plain
\def\@tightleading{1.1}
\def\@doubleleading{1.6}
\def\baselinestretch{\@doubleleading}
\def\tightenlines{\def\baselinestretch{\@tightleading}}
\def\received#1{\gdef\@recvdate{#1}} \received{\relax}
\def\accepted#1{\gdef\@accptdate{#1}} \accepted{\relax}
\def\journalid#1#2{\gdef\@jourvol{#1}\gdef\@jourdate{#2}}
\def\articleid#1#2{\gdef\@startpage{#1}\gdef\@finishpage{#2}}
\def\@rcvaccrule{\vrule\@width1.75in\@height0.5pt\@depth0pt}
\def\dates{{\center\small{\it Received:}\space%
\if\@recvdate\relax\@rcvaccrule\else\@recvdate\fi;%
\hspace{1.5em}{\it Accepted:}\space%
\if\@accptdate\relax\@rcvaccrule\else\@accptdate\fi%
\endcenter}}
\let\ltx@abstract=\abstract
\def\abstract{\dates\ltx@abstract}
\def\title#1{{\def\baselinestretch{\@tightleading}
\center\large\bf#1\endcenter}}
\def\author#1{{\topsep 0pt\center\normalsize#1\endcenter}}
\def\affil#1{\vspace*{-2.5ex}{\topsep 0pt\center\small\it#1\endcenter}}

\skip\footins 4ex plus 1ex minus .5ex
\long\def\@footnotetext#1{\insert\footins{
\def\baselinestretch{\@tightleading}\footnotesize
\interlinepenalty\interfootnotelinepenalty
\splittopskip\footnotesep
\splitmaxdepth \dp\strutbox \floatingpenalty \@MM
\hsize\columnwidth \@parboxrestore
\edef\@currentlabel{\csname p@footnote\endcsname\@thefnmark}\@makefntext
{\rule{\z@}{\footnotesep}\ignorespaces
#1\strut}}}
\long\def\@makefntext#1{\noindent\hbox to\z@{\hss$^{\@thefnmark}$}#1}

\def\tablenotetext#1#2{
\@temptokena={\vspace{.5ex}{\noindent\llap{$^{#1}$}#2}\par}
\@temptokenb=\expandafter{\tblnote@list}
\xdef\tblnote@list{\the\@temptokenb\the\@temptokena}}
\def\spewtablenotes{
\ifx\tblnote@list\@empty
\else
\let\@temptokena=\tblnote@list
\gdef\tblnote@list{\@empty}
\vspace{4.5ex}
\footnoterule
\vspace{.5ex}
{\footnotesize\@temptokena}
\fi}
\newtoks\@temptokenb
\def\tblnote@list{}
\def\endtable{\spewtablenotes\end@float}
\@namedef{endtable*}{\spewtablenotes\end@dblfloat}

\def\@xfloat#1[#2]{\ifhmode \@bsphack\@floatpenalty -\@Mii\else
\@floatpenalty-\@Miii\fi\def\@captype{#1}\ifinner
\@parmoderr\@floatpenalty\z@
\else\@next\@currbox\@freelist{\@tempcnta\csname ftype@#1\endcsname
\multiply\@tempcnta\@xxxii\advance\@tempcnta\sixt@@n
\@tfor \@tempa :=#2\do
{\if\@tempa h\advance\@tempcnta \@ne\fi
\if\@tempa t\advance\@tempcnta \tw@\fi
\if\@tempa b\advance\@tempcnta 4\relax\fi
\if\@tempa p\advance\@tempcnta 8\relax\fi
}\global\count\@currbox\@tempcnta}\@fltovf\fi
\global\setbox\@currbox\vbox\bgroup
\def\baselinestretch{\@tightleading}\@normalsize
\boxmaxdepth\z@
\hsize\columnwidth \@parboxrestore}
\def\@keywordtext{Subject headings}
\def\@keyworddelim{---}
\def\keywords#1{\vspace*{-.7ex}
\if@twocolumn\noindent{\small{\it\@keywordtext:\/}\space\@kwds{#1}}
\else{\quote\small{\it\@keywordtext:\/}\space\@kwds{#1}\endquote}
\fi}

\def\@kwds#1{\def\@kwddlm{}\@for\@kwd:=#1\do
{\@kwddlm\def\@kwddlm{\space\@keyworddelim\penalty\@m\space}{\@kwd}}}
\def\section{\@startsection {section}{1}{\z@}{2.3ex plus 1ex minus
.2ex}{1.5ex plus .2ex}{\normalsize\bf}}
\def\subsection{\@startsection{subsection}{2}{\z@}{2ex plus 1ex minus
.2ex}{1ex plus .2ex}{\normalsize\bf}}
\def\subsubsection{\@startsection{subsubsection}{3}{\z@}{2ex plus
1ex minus .2ex}{1ex plus .2ex}{\normalsize\it}}
\def\acknowledgments{\vskip 3ex plus .8ex minus .4ex}

\def\mathwithsecnums{
\@newctr{equation}[section]
\def\theequation{\hbox{\normalsize\arabic{section}-\arabic{equation}}}}
\def\references{\subsection*{REFERENCES}
\bgroup\parindent=0pt\parskip=\itemsep
\def\refpar{\par\hangindent=3em\hangafter=1}}
\def\endreferences{\refpar\egroup\wgas@pageid}

\def\endthebibliography{\endlist\wgas@pageid}
\def\@biblabel#1{\relax}
\def\@cite#1#2{#1\if@tempswa , #2\fi}
\def\reference{\relax\refpar}
\def\@citex[#1]#2{\if@filesw\immediate\write\@auxout{\string\citation{#2}}\fi
\def\@citea{}\@cite{\@for\@citeb:=#2\do
{\@citea\def\@citea{,\penalty\@m\ }\@ifundefined
{b@\@citeb}{\@warning
{Citation `\@citeb' on page \thepage \space undefined}}%
{\csname b@\@citeb\endcsname}}}{#1}}
\let\jnl@style=\rm
\def\ref@jnl#1{{\jnl@style#1}}
\def\aj{\ref@jnl{AJ}}
\def\araa{\ref@jnl{ARA\&A}}
\def\apj{\ref@jnl{ApJ}}
\def\apjl{\ref@jnl{ApJ}}
\def\apjs{\ref@jnl{ApJS}}
\def\applopt{\ref@jnl{Appl.Optics}}
\def\apss{\ref@jnl{Ap\&SS}}
\def\aap{\ref@jnl{A\&A}}
\def\aapr{\ref@jnl{A\&A~Rev.}}
\def\aaps{\ref@jnl{A\&AS}}
\def\azh{\ref@jnl{AZh}}
\def\baas{\ref@jnl{BAAS}}
\def\jrasc{\ref@jnl{JRASC}}
\def\memras{\ref@jnl{MmRAS}}
\def\mnras{\ref@jnl{MNRAS}}
\def\pra{\ref@jnl{Phys.Rev.A}}
\def\prb{\ref@jnl{Phys.Rev.B}}
\def\prc{\ref@jnl{Phys.Rev.C}}
\def\prd{\ref@jnl{Phys.Rev.D}}
\def\prl{\ref@jnl{Phys.Rev.Lett}}
\def\pasp{\ref@jnl{PASP}}
\def\pasj{\ref@jnl{PASJ}}
\def\qjras{\ref@jnl{QJRAS}}
\def\skytel{\ref@jnl{S\&T}}
\def\solphys{\ref@jnl{Solar~Phys.}}
\def\sovast{\ref@jnl{Soviet~Ast.}}
\def\ssr{\ref@jnl{Space~Sci.Rev.}}
\def\zap{\ref@jnl{ZAp}}

\let\apjsupp=\apjs

\def\la{\mathrel{\hbox{\rlap{\hbox{\lower4pt\hbox{$\sim$}}}\hbox{$<$}}}}
\def\ga{\mathrel{\hbox{\rlap{\hbox{\lower4pt\hbox{$\sim$}}}\hbox{$>$}}}}

\newcount\lecurrentfam
\def\LaTeX{\lecurrentfam=\the\fam \leavevmode L\raise.42ex
\hbox{$\fam\lecurrentfam\scriptstyle\kern-.3em A$}\kern-.15em\TeX}

\makeatother

\def\singlespace{\baselineskip 11.38 pt}
\def\doublespace{\baselineskip 22.76 pt}

\def\mathnew{\mathsurround=0pt}
\def\simov#1#2{\lower .5pt\vbox{\baselineskip0pt \lineskip-.5pt
        \ialign{$\mathnew#1\hfil##\hfil$\crcr#2\crcr\sim\crcr}}}

\def\simless{\mathrel{\mathpalette\simov <}}
\def\eg{{\it e.g.}}
\def\ie{{\it i.e.}}

\def\nature{{\it Nature}}

\def\pCsix{C^{64}}

\def\pCten{C^{1024}}

\def\C{(C)_{\rm max}}
\def\Csix{(C^{64})_{\rm max}}

\def\Cten{(C^{1024})_{\rm max}}

\def\pCbarsix{\bar{C}^{64}}

\def\pCbarten{\bar{C}^{1024}}

\def\Cbar{(\bar{C})_{\rm max}}
\def\Cbarsix{(\bar{C}^{64})_{\rm max}}

\def\Cbarten{(\bar{C}^{1024})_{\rm max}}

\def\Bobs{{\it B}_{\rm obs}}
\def\Vobs{{\it V}_{\rm obs}}
\def\B{{\it B}}
\def\V{{\it V}}

\def\Nb{{N_{\rm bright}}}
\def\Nf{{N_{\rm faint}}}
\def\tdur{t_{\rm dur}}
\def\tsdur{t^{\rm spike}_{\rm dur}}

\def\g{$\gamma$}
\def\etal{{\it et al.}}

\received{1993 March 4}
\accepted{1993 May 24}

\tightenlines
\begin{document}

\title{EVIDENCE FOR TWO DISTINCT MORPHOLOGICAL CLASSES OF GAMMA-RAY
BURSTS FROM THEIR SHORT TIMESCALE VARIABILITY}

\author{D. Q. Lamb, C. Graziani, and I. A. Smith}

\affil{Department of Astronomy and Astrophysics and Enrico Fermi Institute\\
	University of Chicago, Chicago, IL 60637}

\begin{abstract}
We have analyzed the 241 bursts for which peak counts $\C$ exist in the
publicly available Burst and Transient Source Experiment (BATSE)
catalog.  Introducing peak counts in 1024 ms as a measure of burst
brightness $\B$ and the ratio of peak counts in 64 and 1024 ms as a
measure of short timescale variability $\V$, we find a statistically
significant correlation between the brightness and the short timescale
variability of \g-ray bursts.  The bursts which are smoother on short
timescales are both faint and bright, while the bursts which are
variable on short timescales are faint only, suggesting the existence
of two distinct morphological classes of bursts.
\end{abstract}


\section{INTRODUCTION}

The Burst and Transient Source Experiment (BATSE) has detected more
than 600 \g-ray bursts (Meegan \etal\ 1993).  The BATSE catalogue
(Fishman \etal\ 1993) is the largest homogeneous sample of \g-ray
bursts ever assembled and, as such, constitutes a unique resource for
studying burst properties.  To date, however, no clear correlations
between burst properties or distinct morphological classes of bursts
have been found.

The Hertzsprung-Russell (H-R) diagram, in which the brightness of a
star is plotted versus color (the ratio of the flux in two different
energy intervals), proved particularly useful in classifying stars.  A
striking feature of \g-ray bursts is the diversity of their time
histories (see, \eg\, Fishman \etal\ 1992).  Guided by the success of
the H-R diagram for stars and the temporal diversity of \g-ray bursts,
we consider an analogous diagram for \g-ray bursts in which we plot a
measure of burst brightness versus a variability ``color'' (the ratio
of the peak flux or the peak counts in two different time intervals).

We choose $B \equiv \Cbarten$, the expected peak counts in 1024 ms, as
our measure of burst brightness because 1024 ms is the longest time
interval for which peak counts are given in the publicly available
BATSE burst catalogue, and therefore has the best chance of smoothing
out effects due to short timescale variability.  We choose $V \equiv
\Cbarsix / \Cbarten$, the ratio of expected peak counts in 64 and 1024
ms, as our measure of short timescale variability because 64 ms is the
shortest time interval, and 64 and 1024 ms are the most disparate time
intervals, for which peak counts are given in the same catalogue.
There are many other possible measures of variability, and better
measures may be found from detailed analysis of burst time histories
and/or from the acquisition of more data.

Our measure of variability measures an extreme property of the burst
(short timescale variability) at a particular moment during the burst
(the peak of the burst).  Thus it is complementary to global measures
of burst variability, such as power spectra (Kouveliotou \etal\ 1992)
and wavelet analyses (Norris \etal\ 1992, 1993).

\section{ANALYSIS}

The solid lines in Figure 1 show the time histories of two {\it
simulated} \g-ray bursts, defined by $\pCbarsix$, the expected counts
in 64 ms.  Both bursts have durations $\tdur$ = 512 ms; the total
period shown in each time history, including background, is 1024 ms.
For $\pCbarten_{\rm bk}$, the expected background counts in 1024 ms, we
take 2500 counts; this value is typical for the BATSE detector (Meegan
\etal\ 1992).  For $\Cbarten$, the expected peak counts in 1024 ms for
the simulated \g-ray bursts, we choose 500 counts; this value is
typical of the bursts observed by BATSE.  Since $\tdur < 1024$ ms for
both bursts in Figure 1, $B \equiv \Cbarten$ corresponds to the total
number of counts in the burst, and is therefore a rough measure of
burst fluence; for bursts with $\tdur > 1024$ ms, this is no longer
true.  The time history of the first simulated burst has a spike of
duration $\tsdur = 64$ ms, whereas that of the second is flat.  The
expected peak counts in 64 ms of the two bursts are $\Cbarsix = 200$
and 64 counts, respectively.

We desire $B$ and $V$ as our measures of burst brightness and short
timescale variability.  However, BATSE observes $\Bobs \equiv$ $\Cten$
and $\Vobs \equiv$ $\Csix/\Cten$, where $\Csix$ and $\Cten$ are the
peak counts in 64 and 1024 ms, respectively.  The latter differ from
the former because the time history observed by BATSE differs from the
expected time history of the burst due to Poisson fluctuations.  The
dashed lines in Figure 1 are examples of the time histories that BATSE
might actually observe for the two bursts.

Figure 2 (top panel) shows the distribution of burst brightness $\Bobs$
versus short time\-scale variability $\Vobs$ for the 201 bursts in the
publicly available BATSE catalogue for which both $\Csix$ and $\Cten$
exist.\footnotemark \footnotetext{We have omitted from our analysis
four additional bursts (trigger ID numbers 414, 486, 508, and 1346) for
which the $\Csix$ or $\Cten$ listed in the publicly available BATSE
catalogue are erroneous (Howard 1993, private communication).}  The
$\Vobs$-distribution is not independent of $\Bobs$.  In particular,
there are almost no bursts in the upper right quadrant and in a small,
triangular-shaped region in the lower left hand corner of the diagram.

By definition, the $\C$ protocol selects the {\it largest} $\pCsix$ and
$\pCten$ during the burst, irrespective of the other.  Thus $\Csix$
need not come from the time interval corresponding to $\Cten$
(nevertheless, one can show that $1/16 \le \Vobs \le 1$).  Consider for
illustrative purposes the limiting case of a flat burst; \ie, a burst
for which $\Cbarsix = \pCbarsix$ is constant.  If $\tdur \le 64$ ms,
there is only one $\pCsix$ during the burst.   Although $\Csix =
\pCsix$ may differ from $\Cbarsix$ due to Poisson fluctuations, it
equals it, on average.  If $\tdur > 64$ ms, there are several $\pCsix$
during the burst, all of which differ from $\Cbar$ due to Poisson
fluctuations.  Since the $\C$ protocol selects the largest of the
$\pCsix$, it is likely that $\Csix$ exceeds $\Cbarsix$ (see Figure 1).
Thus $\Csix$ is a {\it biased estimator} of $\Cbarsix$ if $\tdur > 64$
ms (Meegan 1993, private communication).  Similarly, $\Cten$ is a
biased estimator of $\Cbarten$.  We must remove this ``Meegan bias''
before evaluating the significance of the pattern in the
($\Bobs,\Vobs$)-diagram.

We have investigated the Meegan bias in detail (Lamb, Boorstein, and
Graziani 1993).  The bias maps the ($\B,\V$)-diagram onto the
($\Bobs,\Vobs$)-diagram.  However, the map is not invertible because of
the Poisson fluctuations of $\Csix$ and $\Cten$ away from $\Cbarsix$
and $\Cbarten$, respectively; rather, it tends to shift $\B$ and $\V$
to larger values of $\Bobs$ and $\Vobs$, and smears them out.  Our
studies show that the bias in $\Vobs$ is zero for a burst whose
duration $\tdur \le 64$ ms, or a burst whose duration $\tdur$ is
arbitrarily long but which has a spike whose duration $\tsdur \le 64$
ms.  The bias in $\Vobs$ increases until $\tdur$ (or $\tsdur$) reaches
1024 ms.  For longer bursts, the bias increases very slowly because the
ratio of the number of $\Csix$ samples to the number of $\Cten$ samples
remains $\approx$ 16.  Thus, the bias for $\tdur$ or $\tsdur = 1024$ ms
closely approximates the maximum bias for any $\tdur$ or $\tsdur$.

We remove the Meegan bias using an approximate method based on a number
of simplifying assumptions.  First, we neglect the bias in $\Bobs
\equiv$ $\Cten$, which is relatively small.  Second, we assume that the
time history of the burst or of the spike during the burst, is flat; in
this limiting case, $\Cbarsix = \pCbarsix$ and $\tdur$ or $\tsdur = 64
\Vobs^{-1}$ ms.  The bias is maximal for this model, as noted earlier;
any difference that remains between the $\V$-distributions for faint
and bright bursts therefore cannot be due to bias.  Third, we adopt the
bias for the minimum of $\tdur$ or $\tsdur = 64 V^{-1}$ ms and 1024 ms,
which closely approximates the maximum bias for any $\tdur$ or
$\tsdur$.  Fourth, we assume that the bias maps a point in the
($\B,\V$)-diagram onto a unique point in the ($\Bobs,\Vobs$)-diagram;
\ie, that the bias map is a $\delta$-function.  This is only
approximately true because of the variances of $\Csix$ and $\Cten$
about $\Cbarsix$ and $\Cbarten$.  These four simplifying assumptions
enable us to invert the bias, and map bursts in the
($\Bobs,\Vobs$)-diagram onto points in the ($\B,\V$)-diagram.

Figure 2 (middle panel) shows contours of constant $\V$ in the ($\Bobs,
\Vobs$)-diagram, taking for the expected background counts in 1024 ms
$\pCbarten_{\rm bk} = 2500$ counts, a value which is typical for the
BATSE detectors.  The bias is zero along the top of the diagram.
Elsewhere, the bias is least in the upper right hand corner and
greatest in the lower left hand corner of the diagram.  This pattern
arises partly from the time variability of the bursts, which is
greatest along the top of the diagram and least along the bottom, and
partly from the relative variances in the total counts $\pCsix_{\rm
bk}+\Csix$ and $\pCten_{\rm bk}+\Cten$, which are least in the upper
right hand corner of the diagram and greatest in the lower left hand
corner.

Bursts that lie below the $\V$ = 1/16 contour (shown as a solid line)
in the lower left hand corner of the ($\Bobs,\Vobs$)-diagram would map
onto points below $\V$ = 1/16 in the ($\B,\V$)-diagram, which is
unphysical.  This is due to our assumption that the burst time history
is flat, for which the bias is maximal, and to our $\delta$-function
approximation for the bias map.   We therefore place a ``floor'' at
$\V$ = 1/16 in the ($\B,\V$)-diagram, and do not allow points mapped
from the $(\Bobs,\Vobs$)-diagram to go below it.

\section{RESULTS}

Figure 2 (bottom panel) shows the resulting distribution of bursts in
the ($\B,\V$)-diagram.  The statistical errors in $B$ and $V$ are a
function only of location in the ($\B,\V$)-diagram; we show them for
six representative locations.  The empty triangular region at the lower
left hand corner of the ($\Bobs,\Vobs$)-diagram has disappeared; it may
therefore be the result of Meegan bias.  In contrast, the presence of
bursts in the upper left hand quadrant and the absence of bursts in the
upper right hand quadrant of the diagram remain.  This means that there
is a lack of bursts that are bright (in 1024 ms) {\it and} have a
bright, short spike (either during the bright 1024 ms or elsewhere
during the burst).

To evaluate the significance of the pattern, we use three $\chi^2$
tests for binned data which address the question of whether or not two
distributions have similar shapes, irrespective of scale, \ie,
irrespective of the total number of objects in each histogram.  The
first is the ``T-test'' of Eadie \etal\ (1971), in which the scales of
the two histograms are set at their maximum likelihood best-fit
values.  The second is the two-histogram test described in {\it
Numerical Recipes} (Press \etal\ 1986), with the histogram having the
smaller number of objects scaled up so that both histograms have the
same number of objects.  The third is an adaptation of the previous
test in which the scale of the histogram with the smaller number of
objects is allowed to vary so as to minimize $\chi^2$

\begin{table}
\doublespace
\centerline{TABLE 1}
\vskip-8pt
$$\vbox{\halign{
\quad#\hfil\quad&\hfil#\hfil\quad&\hfil#\hfil\quad&
\quad\hfil#\hfil\quad&\hfil#\hfil\quad&\hfil#\hfil\quad\cr
\multispan6 \hfil H{\tenrm ISTOGRAMS OF} F{\tenrm AINT}, B{\tenrm
RIGHT}, \hfil\cr
\noalign{\vskip -5pt}
\multispan6 \hfil S{\tenrm MOOTH, AND} V{\tenrm ARIABLE} B{\tenrm
URSTS} \hfil\cr
\noalign{\vskip15pt \hrule \vskip1pt \hrule \smallskip}
&\multispan2{\hfil Two Bin\hfil\quad}&\multispan3{\hfil Three Bin\hfil}\cr
\noalign{\vskip -6pt}
Sample&Bin 1&Bin 2&Bin 1&Bin2&Bin 3\cr
\noalign{\smallskip \hrule \smallskip}
Faint&135&27&126&19&17\cr
\noalign{\vskip -6pt}
Bright&38&1&38&0&1\cr
%
%
Smooth&131&33&81&73&10\cr
\noalign{\vskip -6pt}
Variable&36&1&28&8&1\cr
\noalign{\smallskip \hrule}
}}$$
\parindent=0pt
\singlespace
\end{table}

We divide the bursts into those which are faint ($\log B \le 3.28$) and
bright ($\log B > 3.28$), and construct histograms having two and three
{\it equal} logarithmic bins in variability between $\log V= -1.2$ and
0.  We also divide the bursts into those which are smooth ($\log \V \le
-0.8$) and variable ($\log \V > -0.8$) on short timescales, and
construct histograms having two and three {\it equal} logarithmic bins
in brightness between $\log B$ = 2 and $4.7$.  Table 1 gives the
numbers of bursts in each bin of these histograms, and Table 2 lists
the results of the three tests.\footnotemark \footnotetext{The results
are similar for histograms with more bins.  However, the results are
increasingly suspect because the number of bursts becomes small in many
histogram bins, necessitating the use of Poisson rather than Gaussian
statistics and invalidating the use of the $\chi^2$ distribution for
calculating statistical significance.} The Q-values range from $2.2
\times 10^{-2}$ to $3.5 \times 10^{-6}$, comparing the variability of
faint and bright bursts, and from $1.5 \times 10^{-2}$ to $7.8 \times
10^{-5}$, comparing the brightness of smooth and variable bursts.

\begin{table}

\centerline{TABLE 2}
\vskip-8pt
$$\vbox{\halign{
\quad#\hfil&\quad\hfil#\hfil&\quad\hfil#\hfil&
\quad\hfil#\hfil&\quad\hfil#\hfil&\quad\hfil#\hfil
&\quad\hfil#\hfil&\quad\hfil#\hfil&\quad\hfil#\hfil\cr
\multispan9 \hfil C{\tenrm OMPARISON} B{\tenrm ETWEEN} F{\tenrm AINT}
{\tenrm AND} B{\tenrm RIGHT}, {\tenrm AND} S{\tenrm MOOTH} {\tenrm AND}
V{\tenrm ARIABLE} B{\tenrm URSTS}
\hfil\cr
\noalign{\vskip15pt \hrule \vskip1pt \hrule \smallskip}
&\multispan4 \hfil Faint vs. Bright\hfil&\multispan4 \hfil Smooth vs. Variable
\hfil\cr
%
\hfil Statistical\hfil&\multispan2 \hfil 2 Bins \hfil&\multispan2 \hfil 3 Bins
\hfil&\multispan2 \hfil 2 Bins \hfil&\multispan2 \hfil 3 Bins \hfil \cr
\noalign{\vskip-8pt}
\noalign{\medskip}
\hfil Test\hfil&$\chi^2$&$Q$-value&$\chi^2$&$Q$-value
&$\chi^2$&$Q$-value&$\chi^2$&$Q$-value\cr
\noalign{\smallskip \hrule \smallskip}
T-Test$^a$&$5.2$&$2.2\times 10^{-2}$&$8.3$&$1.6\times 10^{-2}$&
$6.5$&$1.0\times 10^{-2}$&$8.4$&$1.5\times 10^{-2}$\cr
Fixed Scale$^b$&$12$&$4.2\times 10^{-4}$&$25$&$3.5\times 10^{-6}$&
$16$&$4.9\times 10^{-5}$&$10$&$6.4\times 10^{-3}$\cr
Variable Scale$^c$&$12$&$5.1\times 10^{-4}$&$25$&$3.7\times 10^{-6}$&
$16$&$7.8\times 10^{-5}$&$8.9$&$1.2\times 10^{-2}$\cr
\noalign{\smallskip \hrule}
}}$$
\vskip -0.3cm
\tenrm
\noindent $^a$ Two-histogram ``T-test'' (Eadie \etal\ 1971).

\noindent $^b$ Two-histogram test (Press \etal\ 1986).

\noindent $^c$ Adaptation of two-histogram test in Press
\etal\ (1986) (see text).
\end{table}

The disparity between the results of the three tests is due
primarily to the small number of bursts in some histogram bins (see
footnote 2).  We therefore also evaluate the significance of the
correlation in the ($B,V$)-diagram using a statistical test which
compares the mean variabilities of faint and bright bursts, or the
mean brightnesses of smooth and variable bursts, and evaluates the
significance of the differences between the means using a $t$-test
(Press \etal\ 1986).  This test is particularly suitable to the
situation at hand because it implicitly takes into account the
uncertainty in the location of each burst in the ($B,V$)-diagram and,
unlike $\chi^2$ tests, does not require binning the data, which can
necessitate the introduction of Poisson statistics when the number of
bursts is small.

We find that the faint ($\B \le 1900$) bursts have a mean short
timescale variability $\overline{\log V} = -0.97 \pm 0.026$ whereas the
bright ($\B > 1900$) bursts have $\overline{\log V} = -1.05 \pm 0.028$.
Thus the difference of the means divided by their combined variance is
2.2, corresponding to a Q-value of $3.3 \times 10^{-2}$.   We find that
the smooth bursts have a mean brightness $\overline{\log B} = 2.99 \pm
0.036$ whereas the variable bursts have $\overline{\log B} = 2.67 \pm
0.059$.  Thus the difference of the means divided by their combined
variance is 4.5, corresponding to a Q-value of $4.6 \times 10^{-5}$.

We conclude that the difference in short timescale variability between
faint and bright \g-bursts, and the difference in brightness between
bursts that are smooth and variable on short timescales are
significant.

We have checked our results in two ways.  First, we determined the
number of bursts in our study which have gaps in their time histories,
due to malfunctioning of the {\it Compton} Observatory tape recorders.
Using the comments table in the publicly available BATSE catalogue, we
find that 25 of the 48 bursts with data gaps are among the bursts we
have used.  Of these 25, 21 are faint and 4 are bright bursts; three of
the faint bursts and none of the bright bursts are variable on short
timescales.  We conclude that gaps in the time histories of the bursts
cannot account for our results.

Second, we inspected the time histories of the 30 individual bursts out
of the 201 bursts in our sample for which time histories are publicly
available.\footnotemark \footnotetext{Two are in Fishman \etal\ (1992a)
and ten are in Fishman \etal\ (1992b); eighteen others are among the
sample time histories that are publicly available from the Compton
Observatory Science Support Center.} Of these, 17 are faint and 13 are
bright bursts; five of the faint bursts and one of the bright bursts
are variable on short timescales.  In all cases, we were able to verify
from inspection of the time history that the location of the burst in
the ($\B,\V$)-diagram is correct.

\section{DISCUSSION}

The correlation between the brightness $\B$ and the short timescale
variability $\V$ of \g-ray bursts might imply strong source
evolution, or it might reflect the existence of two distinct classes of
\g-ray bursts.  One class might be bursts that exhibit a range of short
timescale variability ($-1.2 \le \log V \le 0$) and are faint, and the
other might be bursts that are smooth on short timescales ($\log V \le
-0.8$) and bright.  Alternatively, one class might be bursts that are
smooth on short timescales ($\log \V \le -0.8$) and range from faint to
bright, and the other might be bursts that are variable on short
timescales ($\log V > -0.8$, corresponding to $\tdur$ or $\tsdur
\simless 0.3$ s) and faint.  Clearly, other decompositions are also
possible.

Using only the correlation between $B$ and $V$ presented here, we
cannot distinguish between these various possibilities.  However, in
subsequent papers, we report statistically significant evidence that
the correlation between $B$ and $V$ reflects the existence of two
distinct morphological classes of \g-ray bursts:  Type I bursts, which
are smooth on short timescales ($\simless$ 0.3 s) and range from faint
to bright, and Type II bursts, which are variable on short timescales
and faint.  Type I bursts also have longer durations and softer spectra
(Lamb and Graziani 1993a), and a flatter brightness distribution (Lamb
and Graziani 1993b) than do Type II bursts.  The dashed lines in the
top and bottom panels of Figure 2 correspond to $\log V = -0.8$, the
cut in variability which separates the two classes.  Because we have
removed the Meegan bias assuming that burst time histories are flat,
for which the bias is maximal, it is likely that we have placed some
bursts which are variable on short timescales in the smooth class, but
not {\it vice-versa}.

\acknowledgments

We gratefully acknowledge the contributions of the scientists who
designed, built, and flew BATSE on the {\it Compton} Observatory, and
whose efforts made possible the work reported here.  We thank Tom
Loredo for many valuable discussions about statistical methodology,
Kevin Hurley for mentioning the potential problem with gaps in the time
history data for individual bursts, and Chip Meegan for pointing out
that peak counts is a biased estimator of expected counts.  We thank
Josh Boorstein for help in investigating the nature of the bias and
Peter Freeman for help in obtaining burst time histories.  CG
acknowledges the support of a NASA Traineeship.  This research was
supported in part by NASA grants NAGW-830, NAGW-1284, NAG5-1454, and
NASW-4690.

\vfill\eject
\begin{figure}
\caption{
Time histories of two {\it simulated} $\gamma$-ray bursts, each lasting
$\tdur$ = 512 ms. (left panel) Burst with a spike in $\pCbarsix$ lasting
$\tsdur$ = 64 ms.  (right panel) Burst with $\pCbarsix$ equal to a
constant.  The solid histograms show $\pCbarsix$, the expected counts
per 64 ms; the dashed histograms show $\pCsix$, the observed counts per
64 ms.  Also labeled are $\Cbarsix$, the expected peak counts per 64
ms, and $\Csix$, the observed peak counts per 64 ms.
}
\end{figure}

\begin{figure}
\caption{
(top panel) Distribution of 201 bursts in the ($\Bobs$,
$\Vobs$)-diagram.  (middle panel) Contours of constant $\V$ in the
($\Bobs$, $\Vobs$)-diagram.  (bottom panel) Distribution of 201 bursts
in the ($\B$, $\V$)-diagram.  The filled and open circles in the
($\Bobs$, $\Vobs$)- and ($\B$, $\V$)-diagrams denote bursts detected by
the 1024 ms trigger and by the 256 or 64 ms triggers, respectively.
The solid lines in the ($\Bobs$, $\Vobs$)- and ($\B$, $\V$)-diagrams
correspond to $\V = 1/16$, the minimum possible variability, while the
dashed lines correspond to $\log \V = -0.8$, the cut in variability
which separates the bursts into two distinct classes.
}
\end{figure}

\end{document}